\documentclass[aps,prl,showpacs,twocolumn,amsmath,amssymb,superscriptaddress]{revtex4}
\usepackage{graphicx}
\usepackage{Jonasmacros}
\usepackage{bm}
\usepackage{fancyheadings}
\begin{document}
\date{\today}
\title{Exchange interaction and Fano resonances in diatomic molecular systems}
\author{J. Fransson}
\email{jonasf@lanl.gov}
\affiliation{Theoretical Division, Los Alamos National Laboratory, Los Alamos, New Mexico 87545, USA}
\affiliation{Center for Nonlinear Studies, Los Alamos National Laboratory, Los Alamos, New Mexico 87545, USA}
\author{A. V. Balatsky}
\email{avb@lanl.gov}
\affiliation{Theoretical Division, Los Alamos National Laboratory, Los Alamos, New Mexico 87545, USA}
\affiliation{Center for Integrated Nanotechnology, Los Alamos National Laboratory, Los Alamos, New Mexico 87545, USA}

\begin{abstract}
We propose a mechanism to use scanning tunneling microscopy (STM) for direct measurements of the two-electron singlet-triplet exchange splitting $J$ in diatomic molecular systems, unsing the coupling between the molecule and the substrate electrons. The different pathways for electrons lead to interference effects and generate kinks in the differential conductance at the energies for the singlet and triplet states. These features are related to Fano resonance due to the branched electron wave functions. The ratio between the tunneling amplitudes through the two atoms can be modulated by spatial movements of the tip along the surface.
\end{abstract}
\maketitle

There are various techniques that allows one to detect and manipulate spin states in the solid state, which attract lots of interest. A partial list include optical detection of electron spin resonance (ESR) in a single molecule \cite{koehler1993}, tunneling through a quantum dot \cite{engel2001}, and, more recently, ESR-scanning tunneling microscopy (ESR-STM) technique \cite{manassen1989,durkan2002}. The interest in ESR-STM is due to the possibility of manipulating single spins \cite{manoharan2002,balatsky2002,koppens2006}, something which is crucial in spintronics and quantum information. Experimentally, modulation in the tunneling current has been observed by STM using spin-unpolarized electron beam \cite{engel2001,manassen1989}. Lately, there has also been a growing interest in using spin-polarized electron beam for direct detection of spin structures \cite{wiesendanger2000}, as well as utilizing the inelastic electron scanning tunneling spectroscopy (IETS) for detection of local spatial variations in electron-boson coupling in molecular systems \cite{madhaven2001,grobis2005}.

Typically in STM measurements with an object located on a substrate surface, the tunneling current can either go directly between the STM tip and the substrate or go via the object. The tunneling electrons are thus branched between different pathways, which gives rise to interference effects when the partial waves merge into one in the tip or the substrate \cite{madhaven2001}. This interference leads to a suppressed transmission probability for the tunneling electrons at certain energies. The suppressed transmission is a fingerprint of Fano resonances \cite{fano1961}, and generally appear in systems where tunneling electrons are branched between different pathways. Recently, Fano resonances have been studied in double and triple quantum dots systems \cite{guevara2003}, where the different pathways are constituted of the different quantum dots.

In this Letter we propose a method to measure the two-electron singlet-triplet (S-T) exchange splitting $J$ in a diatomic molecule by means of STM. The presence of two pathways for the tunneling current between the tip and the substrate, through the diatomic molecule, gives rise to interference effects (Fano resonance) between the electron waves traveling through the singlet and triplet states. In the direct tunneling between the tip and substrate via the molecule, the probability for the tunneling is proportional to $\Gamma_0$. In addition, because of the phase space branching of tunneling possibilities, the tunneling probability is multiplied with the interference probabilitiy $\gamma^2$, hence, the characteristic energy width of the anti-resonance is $\gamma^2\Gamma_0$.
Clearly, the anti-resonances will be measurable in the second derivative of the current whenever $\gamma\ll1$.

\begin{figure}[b]
\begin{center}
\includegraphics[width=8.5cm]{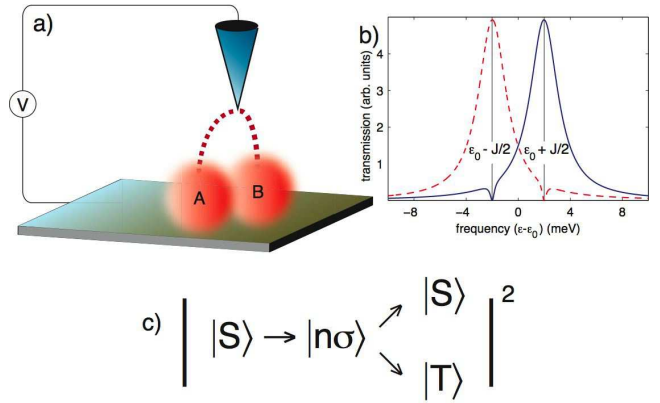}
\end{center}
\caption{(Color online) a) Cartoon of the diatomic $A+B$ molecule coupled to the tip and substrate. The particles interact via exchange interaction $J$. b) Transmission for transitions between the singlet/triplet and the anti-bonding (dashed) and bonding (solid) one-electron states. c) Phase space branching of the tunneling wave functions between the singlet $\ket{S}$ and triplet $\ket{T}$ states. The respective amplitudes of the tunneling through these states have to be added, resulting in the Fano like features.}
\label{fig-system}
\end{figure}
Fano resonances can be realized in a variety of system, ranging from systems with interactions between continuum states and a localized state, to systems where the branching of the wave function through diatomic molecules. In the case we consider here, we also have to include the fact that the one-electron states in a two-level system with the levels being resonant, consist of an anti-bonding and a bonding state (both being spin-degenerate). This modifies the expected transport properties such that transitions between the triplet and the anti-bonding one-electron state generate a dip in the transmission at the energy for this transition, see Fig. \ref{fig-system} b) (dashed). On the other hand, transitions between the singlet and the bonding one-electron state give rise to a dip in the transmission at the energy for this transition, see Fig. \ref{fig-system} b) (solid). In turn, the features that appear at voltages corresponding to both the singlet and triplet states, provide a unique fingerprint which allows the read out of the S-T exchange splitting $J$.

In a regular Fano resonance, interference occur between the different tunneling paths in real space, one path going through the molecule to the substrate whereas the other goes directly into the substrate. Here, on the other hand, the interference occur between different pathways in phase space where the tunneling can occur through a singlet $\ket{S}$ or triplet $\ket{T}$. The respective amplitudes of the tunneling through theses states have to be added, resulting in the Fano like features, see Fig. \ref{fig-system} c).

In general, Fano resonance appearing at voltages that correspond to both the singlet and the triplet states is expected to be found in any model which includes one-electron states and two-electron singlet and triplet configurations, and which accounts for the S-T exchange splitting $J$. For simplicity, however, we will perform our calculations for the interference between the singlet $\ket{S}$ and triplet $\ket{T}=\ket{S=1,\ S_z=0}$. We checked that the interference arising from the other triplet state $\ket{S=1,S_z=\pm1}$ leads to results which only renormalizes the coefficients in the final expressions for the transmission, and therefore can be omitted. Hence, we use a simplified model for the molecule, which both provides the bonding and anti-bonding one-electron states, as well as the two-electron singlet $\ket{S}$ and triplet $\ket{T}=\ket{S=1,\ S_z=0}$ states, i.e.
\begin{equation}
\Hamil_{AB}=\sum_\sigma\dote{0}(\ddagger{A\sigma}\dc{A\sigma}+\ddagger{B\sigma}\dc{B\sigma})
    +J(S_A^+S_B^-+S_A^-S_B^+).
\label{eq-QBmodel}
\end{equation}
Here, $\ddagger{A\sigma/B\sigma}\ (\dc{A\sigma/B\sigma})$ creates (annihilates) an electron in atom $A/B$ at the energy $\dote{0}$, whereas $J$ denotes the exchange splitting, and $S_A^+=\ddagger{A\up}\dc{A\down}$ and $S_A^-=\ddagger{A\down}\dc{A\up}$, and analogously for $B$. In this model, the one-electron states are the (spin-degenerate) anti-bonding and bonding states $\ket{1\sigma}=(\ddagger{A\sigma}-\ddagger{B\sigma})\ket{0}/\sqrt{2}$ and $\ket{2\sigma}=(\ddagger{A\sigma}+\ddagger{B\sigma})\ket{0}/\sqrt{2}$,  respectively, whereas the two-electron states are $\ket{S/T}=[\ddagger{B\down}\ddagger{A\up}\mp\ddagger{B\up}\ddagger{A\down}]\ket{0}/\sqrt{2}$ for the singlet ($S$) and triplet ($T$), respectively.
In order to keep the number of parameters to a minimum, we ignore the splitting between the bonding and anti-bonding one-electron states. This assumption is  not crucial for our subsequent analysis, since the Fano resonance we discuss arise due to interference between the singlet and triplet states, and not between the one-electron states. In order to make contact with experiments, we assume that $J$ corresponds to the exchange splitting which is renormalized by the coupling to the substrate, that is, the  $J$ that will be detected by any measurement in a given setup.

To illustrate the electron interference along different paths consider an electron tunneling from the tip to the substrate via the molecule. Because of the different paths, the electron acquires a different phase depending on whether it travels through $A$ or $B$ and, in addition, whether it makes a transition between the triplet or singlet state and the bonding or anti-bonding one-electron states. To be specific, suppose a spin $\up$ electron leaves the molecule from the singlet state to a medium that hybridizes equally strong to $A$ and $B$, e.g.
\begin{equation}
(\dc{A\up}+\dc{B\up})\ket{S}=
    (-\ddagger{B\down}-\ddagger{A\down})\ket{0}/\sqrt{2}=-\ket{2\down}.
\label{eq-Sa}
\end{equation}
The final state is orthogonal to the anti-bonding state $\ket{1\down}$ Hence, the singlet state couples to the bonding one-electron state. Likewise, supposing that a spin $\up$ electron tunnels out from the triplet state, e.g.
\begin{equation}
(\dc{A\up}+\dc{B\up})\ket{T}=
    (-\ddagger{B\down}+\ddagger{A\down})\ket{0}/\sqrt{2}=\ket{1\down},
\label{eq-Tb}
\end{equation}
shows that the triplet only couples to the anti-bonding one-electron state.
\begin{figure}[b]
\begin{center}
\includegraphics[width=8.5cm]{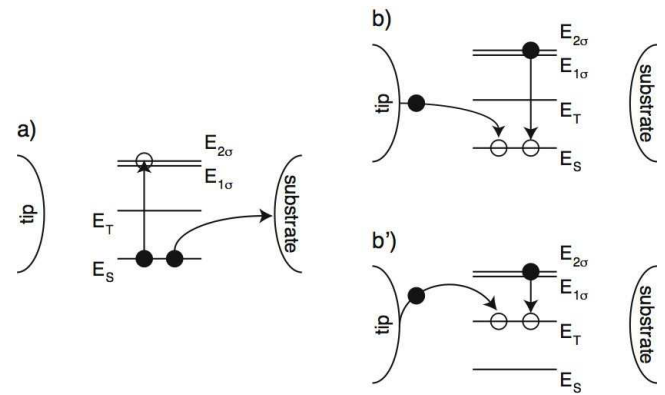}
\end{center}
\caption{Example of tunneling events leading to the interference. First a) an electron leaves the molecule through a transition to the anti-bonding one-electron state $\ket{2\sigma}$. Second, an electron enters the molecule through a transition to either the singlet state b) or to the triplet state b').}
\label{fig-cotunneling}
\end{figure}

The tunneling between the tip/substrate and the molecule is modeled by $\Hamil_T=\sum_{k\sigma}[v_{Ak\sigma}\cdagger{k}\dc{A\sigma}+v_{Bk\sigma}\cdagger{k}\dc{B\sigma}+H.c.]$, where $\cdagger{k}$ creates an electron in the tip/substrate, whereas $v_{Ak\sigma}$ and $v_{Bk\sigma}$ are the tunneling rates between the tip/substrate and  $A$ and $B$, respectively. Reformulating the molecule in terms of its eigenstates, e.g. $\Hamil_{AB}=\sum_{\sigma,n=1,2}E_n\ket{n\sigma}\bra{n\sigma}+E_S\ket{S}\bra{S}+E_T\ket{T}\bra{T}$ with the eigenenergies $E_n=\dote{0}$, $n=1,2$, corresponding to the states $\ket{n\sigma}$, and $E_{S/T}=2\dote{0}\mp J/2$, and neglecting the transitions between the empty and one-electron states, the tunneling Hamiltonian becomes $\Hamil_T=\sum_{k\sigma nX}[v_{k\sigma nX}\cdagger{k}\ket{n\bar\sigma}\bra{X}+H.c.]$, where $X=S,T$, whereas
\begin{equation}
v_{k\sigma nX}=
    v_{Ak\sigma}\bra{n\bar\sigma}\dc{A\sigma}\ket{X}
    +v_{Bk\sigma}\bra{n\bar\sigma}\dc{B\sigma}\ket{X}.
\label{eq-v}
\end{equation}
This form of the tunneling rate reflects the real space branching of the electrons tunneling between the tip and the substrate via the molecule. From this description of the tunneling rate we define the couplings $\Gamma^{L/R}_{nXX'n'}=2\pi\sum_{k\in L/R}v_{k\sigma nX}^*v_{k\sigma n'X'}\delta(\omega-\leade{k})$, where $L/R$ denote states in the tip/substrate. In the present study, the spin currents through the system are identical, thus we do not explicitly indicate the spin in the couplings.

Suppose that the molecule is in the two-electron singlet state. The tunneling current between the tip and the substrate is then mediated by sequences like $\ket{S}\ \rightarrow\ \ket{n\sigma}\ \rightarrow\ \ket{X}$, where $X$ either of the singlet ($S$) or the triplet ($T$) states. Hence, there occur indirect S-T transitions, in the sense that the tunneling may cause a singlet configuration to be turned into a triplet, and vice versa, see Fig. \ref{fig-cotunneling}. These indirect S-T transitions give rise to the interference effects between the different transport channels, which implies that multiple scattering events have to be taken into account in the description of the tunneling through the molecule. These tunneling events leading to the interference resemble cotunneling often discussed in this context \cite{cotunneling}. Here, the multiple scattering events are accounted for through the Green functions (GFs) $G_{nXX'n}(t,0)$ \cite{franssonPRL2002}, where subscripts $nX$ denote the transitions $\ket{n\bar\sigma}\bra{X}$, whereas subscripts $Xn$ denote the conjugate transitions $\ket{X}\bra{n\bar\sigma}$.

We assume, for simplicity, that both atoms couple equally to the substrate and model this by letting $v_{Aq\sigma}=v_{Bq\sigma}=v_{q\sigma}$ for $q\sigma\in R$. From Eqs. (\ref{eq-Tb}) and (\ref{eq-v}) this leads to a vanishing probability for transitions between the triplet and anti-bonding one-electron state, since $|v_{q\sigma 2T}|=|v_{q\sigma}||(\bra{2\bar\sigma}\dc{A\sigma}\ket{T}+\bra{2\bar\sigma}\dc{B\sigma}\ket{T})|=|v_{q\sigma}||-1/2+1/2|=0$. Likewise, from Eqs. (\ref{eq-Sa}) and (\ref{eq-v}), we note that the probability for transitions between the singlet and bonding one-electron state vanishes. Hence, the only non-vanishing coupling matrix elements between the molecule and the substrate are given by $\Gamma^R_{1T}=\Gamma^R_{2S}=\Gamma_0/2$, where $\Gamma_0=2\pi\sum_{k\in L,R}|v_{k\sigma}|^2\delta(\omega-\leade{k})$. As we mentioned, the Fano resonance feature should be present for any bonding and anti-bonding splitting, since the interference we discuss occur between the different tunneling processes $\ket{S}\rightarrow\ket{n\sigma}\rightarrow\ket{S}$ and $\ket{S}\rightarrow\ket{n\sigma}\rightarrow\ket{T}$, and not between the one-electron states, see Fig. \ref{fig-cotunneling}. 

The tunneling between the tip and the molecule, however, is assumed to depend on the spatial position of the tip relatively $A$ and $B$. Introducing the functions $\gamma_{A/B}(|\bfr-\bfr_{A/B}|)$, we model the spatial dependence by putting $v_{Ap\sigma}=\gamma_A(|\bfr-\bfr_A|)v_{p\sigma}$ and $v_{Bp\sigma}=\gamma_B(|\bfr-\bfr_B|)v_{p\sigma}$ for $p\sigma\in L$. The coupling matrix between the anti-bonding one-electron state and the two-electron states for the tunneling between the tip and the molecule then becomes
\begin{eqnarray}
\bfGamma^L_1&=&\left(\begin{array}{ll}
    \Gamma_{1T}^L & \Gamma_{1TS1}^L \\
    \Gamma_{1ST1}^L & \Gamma_{1S}^L
    \end{array}\right)
\nonumber\\&=&
    \frac{\Gamma_0}{8}\left(\begin{array}{ll}
    (\gamma_A+\gamma_B)^2 & \gamma_A^2-\gamma_B^2 \\
    \gamma_A^2-\gamma_B^2 & (\gamma_A-\gamma_B)^2
    \end{array}\right).
\label{eq-GL}
\end{eqnarray}
The coupling matrix between the bonding state and the two-electron states is obtained by letting $\pm\rightarrow\mp$ in the diagonal elements in Eq. (\ref{eq-GL}). The assumption that the atoms couple equally to the substrate is not crucial. Unequal couplings between the atoms the substrate would result in the coupling matrices $\Gamma^R_{1/2}$ being full (structurally equal to $\Gamma^L_{1/2}$). However, since the tip is movable, inequalities in the coupling between the atoms and the substrate can be canceled by relocating the tip to a position where it couples equally to both atoms.

Using the equation of motion technique we find that the Fourier transformed retarded GF, to a good approximation, can be solved by \cite{sc}
\begin{subequations}
\label{eq-GF}
\begin{eqnarray}
G_{nTTn}^r(\omega)&=&\frac{\omega-(E_S-E_n)-\Sigma_{nS}^r(\omega)}
    {C_n(\omega)},
\label{eq-GFT}
\\
G_{nTSn}^r(\omega)&=&\frac{\Sigma_{nTSn}^r(\omega)}{C_n(\omega)},
\label{eq-GFTS}
\\
\Sigma_{nS}^r(\omega)&=&\sum_{k\sigma}
    \frac{|v_{k\sigma nS}|^2}{\omega-\leade{k}+i0^+},
\label{eq-ST}
\\
\Sigma_{nTSn}^r(\omega)&=&\sum_{k\sigma}\frac{v_{k\sigma nT}^*v_{k\sigma nS}}
    {\omega-\leade{k}+i0^+},
\end{eqnarray}
\end{subequations}
where $E_S-E_n=\dote{0}-J/2$. The GFs $G_{nSSn}^r$ and $G_{nSTn}^r$ are obtained from Eq. (\ref{eq-GF}) by letting $T\ (S)\rightarrow S\ (T)$, where $E_T-E_n=\dote{0}+J/2$. In Eq. (\ref{eq-GF}) we have used the function $C_n(\omega)=\det\bfG_n^{r,-1}=(\omega-\omega_{n+})(\omega-\omega_{n-})$, that is the poles of the GF, which are given by
\begin{subequations}
\label{eq-C}
\begin{eqnarray}
\omega_{1\pm}&=&\dote{0}
    -i\frac{\Gamma_0}{8}\left(1+\frac{\gamma_A^2+\gamma_B^2}{2}\right)
\label{eq-C1}\\
&&
    \pm\frac{1}{2}\sqrt{\biggl[J-i\frac{\Gamma_0}{4}(1+\gamma_A\gamma_B)
        \biggr]^2
    -\biggl(\frac{\Gamma_0}{2}\frac{\gamma_A^2-\gamma_B^2}{4}\biggr)^2},
\nonumber\\
\omega_{2\pm}&=&\dote{0}
    -i\frac{\Gamma_0}{8}\left(1+\frac{\gamma_A^2+\gamma_B^2}{2}\right)
\label{eq-C2}
\\&&
    \pm\frac{1}{2}\sqrt{\biggl[J+i\frac{\Gamma_0}{4}(1+\gamma_A\gamma_B)
        \biggr]^2
    -\biggl(\frac{\Gamma_0}{2}\frac{\gamma_A^2-\gamma_B^2}{4}\biggr)^2}.
\nonumber
\end{eqnarray}
\end{subequations}
These expressions are obtained by neglecting the real parts of the self-energies $\Sigma^r_{nS},\ \Sigma_{nTSn}^r$, we have ${\bf \Sigma}^r_{1(2)}(\omega)=-i(\bfGamma^L_{1(2)}+\bfGamma^R_{1(2)})/2$. Here, subscript $1\ (2)$ refers to the anti-bonding (bonding) one-electron state, whereas $+\ (-)$ refers to the triplet (singlet) two-electron state. Hence, the energy $\omega_{1+}$ corresponds to transitions between the triplet and anti-bonding states.

From Eqs. (\ref{eq-C}) we extract the widths of the transitions between the one- and two-electron states. Assuming $\gamma_B=\kappa\gamma_A$, where $\gamma_A\ll1$ and $0\leq\kappa\leq1$, we find that the widths of the transitions between the triplet (singlet) and bonding (anti-bonding) states are significantly smaller than the transitions between the triplet (singlet) and the anti-bonding (bonding) states. Indeed, while the widths of the main peaks at $\omega_{1+}$ and $\omega_{2-}$ are approximately $\Gamma_0/4$, the widths of the narrow peaks at $\omega_{1-}$ and $\omega_{2+}$ are approximately 
$(\Gamma_0/4)(1+\kappa^2)\gamma_A^2/4$.
Hence, for small $\gamma_A$ we expect sharp features in the differential conductance for the system. However, due of the small width of the dips, these features will be easier to observe in the derivative of the differential conductance, i.e. the second derivative of the current $I\sim\tr\int{\cal T}(\omega)[f_L(\omega)-f_R(\omega)]d\omega$ which here is given by
\begin{eqnarray}
\frac{\partial^2I}{\partial V^2}&\sim&
    \int{\cal T}(\omega)\biggl(
        \frac{\tanh{[\beta(\omega-\mu_L)/2]}}{\cosh^2{[\beta(\omega-\mu_L)/2]}}
\nonumber\\&&\hspace{2cm}
        -\frac{\tanh{[\beta(\omega-\mu_R)/2]}}{\cosh^2{[\beta(\omega-\mu_R)/2]}}
        \biggr)d\omega,
\label{eq-d2J}
\end{eqnarray}
where ${\cal T}_n(\omega)=\tr\bfGamma^L_n\bfG^r_n(\omega)\bfGamma^R_n\bfG^a_n(\omega)$ is the transmission coefficient \cite{meir1992}, whereas $\mu_{L/R}$ is the chemical potential of the tip/substrate. Under the given conditions we find that the total transmission ${\cal T}=\sum_n{\cal T}_n$ through the molecule is given by
\begin{eqnarray}
{\cal T}(\omega)&=&\left(\frac{\Gamma_0}{4}\right)^2
    (\gamma_A+\gamma_B)^2
\nonumber\\&&
    \times\biggl(
        \left|\frac{\omega-\dote{0}+J/2}{C_1(\omega)}\right|^2
        +\left|\frac{\omega-\dote{0}-J/2}{C_2(\omega)}\right|^2
        \biggr),
\label{eq-T}
\end{eqnarray}
which shows dips at the energies corresponding to the transitions between the one-electron and two-electron singlet and triplet states, e.g. at $\omega=\dote{0}\mp J/2$, respectively, independent of $\gamma_{A/B}$. From the above analysis of the widths, these dips are expected to be sharp since their positions in the transmission exactly correspond to the roots $\omega_{1-}$ and $\omega_{2+}$, respectively, for small $\gamma_A$.

Asymmetric coupling between both the tip and substrate and the atoms generate a shift in the positions of the transmission dips \cite{genT}. This result shows that the distance between the transmission dips deviates from $J$ by at most 6 \% for asymmetries $v_{Ap(q)\sigma}/v_{Bp(q)\sigma}=\gamma_{L(R)}\leq0.7$ \cite{genT}. Thus, even for asymmetric coupling, this will still be a good measure of $J$.

In conclusion we propose to use the interference between different paths of the tunneling electron wavefunction as a tool to detect the S-T splitting in a single molecule.  This splitting could be observed as features in the tunneling conductance. Becasue the features we find are rather narrow they can be, more prominently, revealed in the $\partial^2I/\partial V^2$ peaks that would correspond to the S-T splitting. The splitting produces interference between the possible pathways for electrons to tunnel through virtual states with different energies, identical to the classical Fano resonance arguments \cite{fano1961}. This effect is similar to IETS features seen in inelastic scattering off non-magnetic and magnetic excitations in molecules \cite{grobis2005}. With current expertimental capabilties in STM one can easily address single diatomic molecules with either spin-polarized STM \cite{wiesendanger2000} or with position dependent IETS measurements \cite{grobis2005}.

This work has been supported by DOE BES and by LDRD at Los Alamos National Laboratory. We are grateful A. Kubetzka, M. Bode, R. Wiesendanger for useful discussions.


\begin{thebibliography}{20}
\bibitem{koehler1993} J. K\"ohler, J. A. J. M. Disselhorst, M. C. J. M. Donckers, E. J. J. Groenen, J. Schmidt, and W. E. Moerner, Nature {\bf 363}, 242 (1993); J. Wrachtrup, C. von Borczyskowski, J. Bernard, M. Orrit, and R. Brown, \emph{ibid}. {\bf 363}, 244 (1993); Phys. Rev. Lett. {\bf 71}, 3565 (1993).
\bibitem{engel2001} H. -A. Engel and D. Loss, Phys. Rev. Lett. {\bf 86}, 4648 (2001); Phys. Rev. B, {\bf 65}, 195321 (2002).
\bibitem{manassen1989} Y. Manassen, R. J. Hamers, J. E. Demuth, and A. J. Castellano, Jr., Phys. Rev. Lett. {\bf 62}, 2531 (1989); D. Shachal and Y. Manassen, Phys. Rev. B, {\bf 46}, 4795 (1992); Y. Manassen, J. Magn. Reson. {\bf 126}, 133, (1997); Y. Manassen, I. Mukhopadhyay, and N. R. Rao, Phys. Rev. B, {\bf 61}, 16223 (2000).
\bibitem{durkan2002} C. Durkan and M. E. Welland, Apply. Phys. Lett. {\bf 80}, 458 (2002).
\bibitem{manoharan2002} H. Manoharan, Nature {\bf 416}, 24 (2002); H. Manoharan, C. P. Lutz, and D. Eigler, Nature {\bf 403}, 512 (2000).
\bibitem{balatsky2002} A. V. Balatsky and I. Martin, Quantum Inf. Process. {\bf 1}, 53 (2002); A. V. Balatsky, Y. Manassen, and R. Salem, Phys. Rev. B, {\bf 66}, 195416 (2002).
\bibitem{koppens2006} F. H. L. Koppens, C. Buizert, K. J. Tielrooij, I. T. Vink, K. C. Nowack, T. Meunier, L. P. Kouwenhoven, and L. M. K. Vandersypen, Nature {\bf 442}, 766 (2006).
\bibitem{wiesendanger2000} S. Heinze, . Bode, A. Kubetzka, O. Pietzsch, X. Nie, S. Bl\"ugel, and R. Wiesendanger, Science {\bf 288}, 1805 (2000); A. Kubetzka , M. Bode, O. Pietzsch, and R. Wiesendanger, Phys. Rev. Lett. {\bf 88}, 057201 (2002); A. Wachowiak \etal, Science {\bf 298}, 577 (2002); J. Wiebe, A. Wachowiak, F. Meier, D. Haude, T. Foster, M. Morgenstern, and R. Wiesendanger, Rev. Sci. Instrum. {\bf 75}, 4871 (2004).
\bibitem{madhaven2001} V. Madhavan, W. Chen, T. Jamneala, M. F. Crommie, and N. S. Wingreen, Phys. Rev. B, {\bf 64}, 165412 (2001).
\bibitem{grobis2005} M. Grobis, K. H. Khoo, R. Yamachika, X. Lu, K. Nagaoka, S. G. Louie, M. F. Crommie, H. Kato, and H. Shinohara, Phys. Rev. Lett. {\bf 94}, 136802 (2005).
\bibitem{fano1961} U. Fano, Phys. Rev. {\bf 124}, 1866 (1961).
\bibitem{guevara2003} M. L. Ladr\'on de Guevara, F. Claro, and P. A. Orellana, Phys. Rev. B, {\bf 67}, 195335; P. A. Orellana, M. L. Ladr\'on de Guevara, and F. Claro \emph{ibid}, {\bf 70}, 233315 (2004); M. L. Ladr\'on de Guevara and P. A. Orellana, \emph{ibid}, {\bf 73}, 205303 (2006).
\bibitem{cotunneling} D. V. Averin and A. A. Odinstov, Physics Letters A, {\bf 140}, 251 (1989); D. V. Averin and Y. V. Nazarov, Phys. Rev. Lett. {\bf 65}, 2446 (1990).
\bibitem{franssonPRL2002} J. Fransson, O. Eriksson, and I. Sandalov, Phys. Rev. Lett. {\bf 88}, 226601 (2002); J. Fransson, Phys. Rev. B, {\bf 72}, 075314 (2005).
\bibitem{sc} These GFs should be calculated self-consistently with respect to the population numbers of the various many-body states \cite{franssonPRL2002}. However, this procedure is omitted here for simplicity since the self-consistent treatment of the GFs only changes quantitative details concerning the transport properties, not the main qualitative features.
\bibitem{meir1992} Y. Meir and N. S. Wingreen, Phys. Rev. Lett. {\bf 68}, 2512 (1992).
\bibitem{genT} The asymmetries $v_{Ap(q)\sigma}=\gamma_{L(R)}v_{Bp(q)\sigma}$ results in ${\cal T}(\omega)=2(1+\gamma_L\gamma_R)(\Gamma_0/8)^2(|(\omega-\dote{0}+qJ/2)/C_1(\omega)|^2+|(\omega-\dote{0}-qJ/2)/C_2(\omega)|^2)$,
with $q=(\gamma_L+\gamma_R)/(1+\gamma_L\gamma_R)$ Assuming $\gamma_{L(R)}\sim0.7$ we obtain 6 \% shift of the transmission dips as quoted in the text.
\end{thebibliography}
\end{document}